\documentclass[aps,pra,twocolumn,superscriptaddress,raggedbottom,showpacs,floatfix,longbibliography]{revtex4-1}
\usepackage{amsmath,amsfonts,amssymb,amsthm,graphics}
\usepackage[next]{inputenc}
\usepackage[colorlinks=true,citecolor=blue,linkcolor=blue]{hyperref}
\usepackage{bbm}
\usepackage{bbold}
\usepackage{booktabs}
\usepackage{multirow}
\usepackage{hhline}
\usepackage{comment}
\usepackage{float}
\usepackage{latexsym}
\usepackage[dvipsnames]{xcolor}
\usepackage{graphicx}
\usepackage{epstopdf}
\usepackage{bm}% bold math
\usepackage{verbatim}
\usepackage{siunitx}

\newcommand{\mc}{\mathcal}

\newcommand{\s}{s}  % spin index of condensate
\newcommand{\p}{\varphi}  % coupling strength
\newcommand{\dt}{dt}  % time between measurements
\renewcommand{\d}{d}  % stochastic increment 
\newcommand{\up}{\uparrow}  % order parameter index
\newcommand{\down}{\downarrow}  % other order parameter index
\newcommand{\e}{\varepsilon}  % estimator symbol
\newcommand{\ramprate}{\gamma}  % ramp off rate for cooling 

\begin{document}

\title{Feedback Induced Magnetic Phases in Binary Bose-Einstein Condensates}
\author{Hilary M. Hurst}
\affiliation{Joint Quantum Institute, National Institute of Standards and Technology, and University of Maryland, Gaithersburg, Maryland, 20899, USA}
\affiliation{Department of Physics and Astronomy, San Jos\'{e} State University, San Jos\'{e}, California, 95192, USA}
\author{Shangjie Guo}
\affiliation{Joint Quantum Institute and Department of Physics, University of Maryland, College Park, Maryland 20742, USA}
\author{I. B. Spielman}
\affiliation{Joint Quantum Institute, National Institute of Standards and Technology, and University of Maryland, Gaithersburg, Maryland, 20899, USA}

\begin{abstract}
Weak measurement in tandem with real-time feedback control is a new route toward engineering novel non-equilibrium quantum matter. Here we develop a theoretical toolbox for quantum feedback control of multicomponent Bose-Einstein condensates (BECs) using backaction-limited weak measurements in conjunction with spatially resolved feedback. Feedback in the form of a single-particle potential can introduce effective interactions that enter into the stochastic equation governing system dynamics. The effective interactions are tunable and can be made analogous to Feshbach resonances -- spin-independent and spin-dependent -- but without changing atomic scattering parameters. Feedback cooling prevents runaway heating due to measurement backaction and we present an analytical model to explain its effectiveness. We showcase our toolbox by studying a two-component BEC using a stochastic mean-field theory, where feedback induces a phase transition between easy-axis ferromagnet and spin-disordered paramagnet phases. We present the steady-state phase diagram as a function of intrinsic and effective spin-dependent interaction strengths. Our result demonstrates that closed-loop quantum control of Bose-Einstein condensates is a powerful new tool for quantum engineering in cold-atom systems. 
\end{abstract}

\maketitle

\section{Introduction}

% !TEX root = main.tex
% Informs TeXShop to look for the main file. 

Quantum gas experiments have exquisite control over the low-energy Hamiltonian governing system dynamics, providing demonstrated opportunities to study interacting many-body quantum systems with great precision. As a result, ultracold atoms have emerged as a leading platform in `analog quantum simulation'~\cite{Bloch2008,  Cirac2012, Georgescu2014, Hodgman2011, Gross2017, Zache2020}, where experiments have successfully explored condensed-matter phenomena such as the superfluid-Mott insulator transition~\cite{Greiner2002}, the BEC-BCS crossover~\cite{Bertenstein2004, Bourdel2004}, and spin-orbit coupling~\cite{Lin2011}. Cutting-edge experiments now realize systems with long-range interactions~\cite{Landig2016} or novel non-equilibrium dynamics~\cite{Ronzheimer2013, Kohlert2019}. In contrast, quantum simulation of open systems remains relatively unexplored~\cite{Solano2019}, and careful application of feedback control to many-body quantum systems is a new approach toward this goal.

Feedback control of many-body systems could enable observation of a wide range of new phenomena in dynamical steady state, where a potentially larger class of states are possible than in thermal equilibrium~\cite{Polkovnikov2011, Heyl2018}. Existing proposals include preparation of many-body pure states via reservoir engineering~\cite{Diehl2008, Kraus2008, Verstraete2009, Laflamme2017}, nonthermal steady states~\cite{Rigol2009, Abanin2019}, stable non-Abelian vortices~\cite{Mawson2019}, or time crystals~\cite{Zhang2017b}. Here, we showcase the flexibility of weak measurements coupled with spatially resolved feedback for quantum simulation of time-dependent effective Hamiltonians using a two-component Bose-Einstein condensate (BEC) as a model spinor system~\cite{Trippenbach2000, Kawaguchi2012, Stamper2013}.

We develop a theory of weak measurement and classical feedback in weakly interacting quantum systems framed in the context of quantum control theory~\cite{Zhang2017}. Using our general formalism we investigate the steady-state phases of a two-component BEC subject to weak measurement and classical feedback via a spin dependent applied potential, enabling both density and spin dependent feedback protocols.

Spatially local feedback can result in spin-dependent effective interaction terms in the stochastic equation governing condensate dynamics. Depending on the interplay of intrinsic and effective (i.e. feedback-induced) spin-dependent interactions, the condensate steady-state phase is either an easy-axis ferromagnet or spin-disordered paramagnet. The effective interaction is tunable via the gain of the feedback signal, enabling a reversible, feedback-induced phase transition. The transition is reminiscent of what is achieved by tuning intrinsic interactions via a spin-dependent Feshbach resonance~\cite{Theis2004}, however here the atomic scattering lengths remain unchanged. We develop a signal filtering and cooling scheme to minimize heating and show that the condensate remains intact under feedback and measurement backaction. Our result opens the door to engineering new dynamical and/or spatially dependent effective interactions in quantum gases via closed-loop feedback control. 

Previous works have considered quantum control protocols for BECs~\cite{Haine2004, Wilson2007, Szigeti2009, Szigeti2010, Hush2013, Wade2015, Ilo2014, Wang2016, Wade2016}. Feedback schemes thus far presented have focused on driving a condensate to it's ground state by altering the position and strength of a harmonic trapping potential~\cite{Haine2004, Wilson2007, Szigeti2009, Szigeti2010, Hush2013}, or to deteministically prepare a target state~\cite{Wade2015, Wade2016}, possibly for quantum memory applications~\cite{Ilo2014, Wang2016}. Here we move beyond the realm of specific state control toward implementation of designer effective Hamiltonians or Louivillians with possibly unknown dynamical steady states. 

The paper is structured as follows: In Sec.~\ref{Sec:summary} we present our main formal results, including the stochastic equation describing condensate dynamics, and introduce a toy model illustrating the salient features of the control protocol. We show that locally applied feedback induces a phase transition between easy-axis ferromagnetic and disordered paramagnetic phases in a two-component condensate.

In Sec.~\ref{Sec:cooling} we elaborate on our feedback cooling protocol and characterize the resulting steady state via condensate fraction, Von Neumann entropy, and energy. We show that heating due to measurement backaction can be effectively mitigated by feedback cooling. In Sec.~\ref{Sec:spinfeedback} we discuss the feedback-induced steady-state phases in more detail and elucidate the nature of the phase transition in our system. We conclude in Sec.~\ref{Sec:conclusion}.

\section{Summary of Results~\label{Sec:summary}}

% !TEX root = main.tex
% Informs TeXShop to look for the main file. 

\subsection{General Formalism~\label{SubSec:Formalism}}
We model dispersive imaging of a quasi-one-dimensional (1D) multicomponent Bose-Einstein condensate of length $L$ via spin resolved phase-contrast imaging~\cite{Andrews1997} and we label individual components by an index $\s$. We consider time and space resolved measurements of atomic density $\hat{n}_\s(x, t)$ in each component using the Gaussian measurement model developed in detail in Ref.~\cite{Hurst2019}. Stroboscopic weak measurements with strength $\p$ result in the measurement signal
\begin{equation}
\mc{M}_\s(x, t) = \langle\hat{n}_{\s}(x, t)\rangle + \frac{m_\s(x)}{\p}, 
\label{Eqn:Mresult}
\end{equation}
where $m_{\s}(x)$ describes spatiotemporal quantum projection noise associated with the measurement. The measurement is characterized by Fourier domain Gaussian statistics $\overline{\tilde{m}_{\s,k}} = 0$ and $\overline{\tilde{m}_{\s,k}\tilde{m}_{\s,k'}} = L\Theta(|k|-k_{\rm c})\overline{dW_{\s,k}dW_{\s',k'}}/2\dt^2$, where $dW_{\s,k}$ is a Wiener increment with $\overline{dW_{\s,k}} = 0$ and $\overline{dW_{\s,k}dW_{\s',k'}} = \dt\delta_{\s{\s}'}\delta_{kk'}$ for a time increment $\dt$~\cite{anote}. The Heaviside function $\Theta$ enforces a momentum cutoff at $k_{\rm c} = 2\pi/\lambda$, accounting for the fact that the physical measurement process can only resolve information with length scales larger than $\lambda/2\pi$. The observer does not directly obtain information about the condensate phase using this protocol.

We use the aggregate measurement result $\mc{M}$, a function of $x$ and $\s$, to generate feedback signals in the form of a single-particle potential $\check{V}\left[\mc{M}\right]$, where $\check{\cdot}$ indicates an operator in component space. In this work we consider a potential which is local in space. 

We describe the condensate in the mean-field approximation using a complex spinor order parameter $\Psi(x) = (\psi_1(x), \psi_2(x), \ldots)^T$, where $\psi_\s(x)$ is a classical field describing the dynamics of component $\s$. The total density is $n(x) = \Psi^\dagger(x)\check{\mathbb{1}}\Psi(x)$ and the order parameter is normalized to the number of particles, $N = \int dx~n(x)$. From Eq.~\eqref{Eqn:Mresult} the measurement results at the mean-field level therefore depend on the field amplitude via $\langle\hat{n}_{\s}(x)\rangle \rightarrow |\psi_\s(x)|^2$. Measurement backaction leads to stochastic evolution of the order parameter, which results in condensate heating~\cite{Dalvit2002, Hurst2019} in the absence of a cooling protocol, which we describe in Sec.~\ref{Sec:cooling}.

The combined measurement and quantum control process is described by a stochastic equation of motion
\begin{equation}
\d\Psi(x) = \left.\d\Psi(x)\right|_{\rm H} + \left.\d\Psi(x)\right|_{\rm M} + \left.\d\Psi(x)\right|_{\rm F}, 
\end{equation}
for the condensate order parameter $\Psi(x)$. Here
\begin{align}
\left.\d\psi_{\s}(x)\right|_{\rm H} &= -\frac{i}{\hbar}\left[\hat{\mc{H}}_{\s{\s}'}(x) - \mu\delta_{\s{\s}'}\right]\psi_{{\s}'}(x) \dt,\label{Eqn:dpsiH}\\
\left.\d\psi_{\s}(x)\right|_{\rm M} &= \left[-\frac{\p^2k_{\rm c}}{4\pi} + \p m_{\s}(x)\right]\psi_{\s}(x)\dt\label{Eqn:dpsiM},\\  
\left.\d\psi_{\s}(x)\right|_{\rm F} &= -\frac{i}{\hbar}V_{\s{\s}'}[\mc{M}](x)\psi_{{\s}'}(x) \dt, \label{Eqn:dpsif}
\end{align}
denote contributions from unitary (i.e. closed system) evolution, measurement backaction, and feedback, respectively and $\mu$ is the chemical potential. We adopt the implied summation convention over repeated indices and set $\hbar = 1$. 

Using this general formalism we study a condensate of $^{87}$Rb atoms from which we select two hyperfine states, yeilding a two-component condensate~\cite{Stamper2013, De2014} with components denoted by $\s =$ $\up, \down$. The Hamiltonian in Eq.~\eqref{Eqn:dpsiH} is the usual Gross-Pitaevskii equation (GPE) describing closed system dynamics, which takes the explicit form 
\begin{equation}
\hat{\mathcal{H}}_{s{\s}'}\psi_{{\s}'}= \left[\hat{H}_0  + u_0n\right]\mathbb{1}_{\s{\s}'}\psi_{{\s}'} + u_2S_z\sigma^z_{\s{\s}'}\psi_{{\s}'},
\label{Eqn:H0}
\end{equation}
for two component condensates, with $(x,t)$ indices suppressed for clarity. Here, $S_z(x) = \Psi^\dagger(x)\check{\sigma}^z\Psi(x)$ indicates the spin density and $\check{\boldsymbol{\sigma}} = (\check{\sigma}^x, \check{\sigma}^y, \check{\sigma}^z)$ is a vector of the Pauli operators. The single particle Hamiltonian is $\hat{H}_0 = \hat{p}^2/2m_{\rm a}$ for atoms of mass $m_{\rm a}$. The intrinsic spin-independent $u_0$ and spin-dependent $u_2$ interaction strengths serve to define $\xi = 1/\sqrt{2m_{\rm a}\mu}$ and $\xi_{\rm s} = \xi \sqrt{u_0/2|u_2|}$, the healing length and spin-healing length respectively.

Equation~\eqref{Eqn:dpsiM} describes measurement backaction. Separate measurements of each condensate component result in independent backaction noise $m_{\s}(x)$. Equation~\eqref{Eqn:dpsif} describes feedback, applied via the potential term $\check{V}\left[\mc{M}\right]$. The feedback potential combines a deterministic part containing information about the condensate dynamics with a stochastic part due quantum projection noise. 
Therefore, both $\left.\d\Psi\right|_{\rm F}$ and $\left.\d\Psi\right|_{\rm M}$ contribute to stochastic condensate dynamics. When each individual measurement is very weak, the density of noncondensed particles remains low. Therefore we assume $\Psi(x)$ to be well described by a lowest order Hartree-Fock theory throughout it's evolution. This assumption is validated in Sec.~\ref{SubSec:BdG} and~\ref{SubSec:continuouscooling}. 

\subsection{Key Feedback Concepts}

Our aim is to develop feedback schemes which add new effective interaction terms to the Hamiltonian while minimizing quantum projection noise. We illustrate the core concept of feedback using a toy model. The toy model is a simplified version of the feedback protocols developed in later sections, that nonetheless illustrates a key result: weak measurements combined with feedback can be used to engineer new effective Hamiltonians. 

\subsubsection{Toy Model \label{SubSubSec:toy}}
Here we construct a minimal model of measurement and feedback for single component systems, and therefore suppress the component index $\s$. We weakly measure the density, then apply a proportional feedback potential
\begin{equation}
V\left[\mathcal{M}\right](x, t) = g_0\mathcal{M}(x, t), 
\label{Eqn:spinlessToymodelV}
\end{equation}
where the gain parameter $g_0$ denotes the feedback strength. Inserting Eq.~\eqref{Eqn:Mresult} into Eq.~\eqref{Eqn:spinlessToymodelV} gives a feedback potential with two contributions. The first is an effective mean-field interaction
\begin{equation}
V^{\rm eff}(x, t) = g_0n(x, t), 
\end{equation}
and the second is a stochastic contribution 
\begin{equation}
V^{\rm fluct}(x, t)  = \frac{g_0m(x)}{\p}.
\label{Eqn:Vfluct}
\end{equation}
By direct substitution of $V\left[\mathcal{M}\right]$ into Eq.~\eqref{Eqn:dpsif}, the dynamical Eqs.~\eqref{Eqn:dpsiH}-\eqref{Eqn:dpsif} reduce to two equations $\d\Psi(x) = \left.\d\Psi(x)\right|_{\rm H'} + \left.\d\Psi(x)\right|_{\rm M'}$ with modified unitary evolution and stochastic terms,  
\begin{align}
\left.\d\psi(x)\right|_{\rm H'} &= -i\left[\hat{\mc{H}}^{\rm eff}(x) - \mu\right]\psi(x)dt \label{Eqn:dpsiHprime} \\
\left.\d\psi(x)\right|_{\rm M'} &=\left[-\frac{\p^2k_{\rm c}}{4\pi} + \left(\p-i\frac{g_0}{\p}\right)m(x)\right]\psi(x)\dt\label{Eqn:dpsiMprime}.
\end{align}
The effective Hamiltonian $\hat{\mc{H}}^{\rm eff}(x)$ has the same form as the spin-independent term in Eq.~\eqref{Eqn:H0}, but with $u_0$ replaced by an effective interaction constant $u^{\rm eff}_0 = u_0 + g_0$. Likewise, the noise in the stochastic evolution is modified due to the contribution of $V^{\rm fluct}(x, t)$. This simple model illustrates how feedback can be used to create new effective Hamiltonians with modified interaction terms. 

Returning to the two-component case, we consider the spin-dependent feedback potential 
\begin{equation}
\check{V}[\mc{M}](x, t) = g_0\mc{M}_n(x, t)\check{\mathbb{1}}+ g_2\mc{M}_z(x, t)\check{\sigma}^z,
\label{Eqn:Vexample}
\end{equation}
describing separate contributions to the density and spin sectors controlled by independent gain parameters $g_0$ and $g_2$, respectively. Measurement signals $\mathcal{M}_\s$ are used to calculate total density and spin density, given by $\mathcal{M}_n = \mathcal{M}_\up + \mathcal{M}_\down$ and $\mathcal{M}_z = \mathcal{M}_\up - \mathcal{M}_\down$, respectively. Following the same algebraic arguments, the feedback potential~\eqref{Eqn:Vexample} leads to effective interaction strengths $u^{\rm eff}_0 = u_0 + g_0$, $u^{\rm eff}_2 = u_2 + g_2$, along with modified stochastic noise on each component $\psi_\s$.

In the following, we use this guiding prinicple to develop a measurement and feedback scheme which controls the magnetic properties of a two-component condensate without changing the internal interaction parameters. The simplified protocol presented in this section is impractical due to runaway heating~\cite{Hurst2019}, from the repeated and uncompensated application of the stochastic potential in Eq.~\eqref{Eqn:dpsiMprime}. In Sec.~\ref{Sec:cooling} we introduce a feedback cooling protocol that prevents runaway heating and thus completes our toolbox for quantum feedback control.

\subsubsection{Signal Filtering}
In the toy model above, the feedback potential is governed only by local in time measurement results. Because Eqs.~\eqref{Eqn:dpsiH}-\eqref{Eqn:dpsif} describe continuous time evolution, the effect of $V^{\rm fluct}(x, t)$ in Eq.~\eqref{Eqn:Vfluct} would seem to diverge as $\dt \rightarrow 0$. However, any measurement signal $\mathcal{M}_i(x,t)$ can be filtered in time to provide a running best estimate of the measured observable $i$ (where $i = n, z,$ etc.).

The resulting estimator $\e_i$ is derived from $\mathcal{M}_i$ via the low-pass filter
\begin{equation}
\tau_i\dot{\e}_i(x, t) + \e_i(x, t)  = \mc{M}_i(x,t), 
\label{Eqn:LowPass1}
\end{equation}
i.e. 
\begin{equation}
\e_i(x, t) = \frac{1}{\tau_i}\int_{-\infty}^t dt'~\mc{M}_i(x,t')e^{-(t-t')/\tau_i}, 
\label{Eqn:LowPassintegrated}
\end{equation}
where $\tau_i$ is the filter time constant and $\mc{M}_i(x,t)$ indicates the unfiltered measurement signal. This process filters the contribution of projection noise present at timescales below $\tau_i$, making $\tau_i$ the effective measurement time associated with the estimator $\e_i$.

We derive all of our feedback potentials using estimators $\varepsilon_i$ instead of measurement signals $\mc{M}_i$, thereby controlling the noise applied to the system via feedback. In our feedback scheme we use separate estimators of the total density, spin density, or density in component $\s$, denoted $\e_n$, $\e_{z}$, $\e_{\s}$, respectively, which can have different filter time constants $\tau_n$, $\tau_z$, and $\tau_\s$.

\subsection{Feedback Induced Magnetic Phases \label{SubSec:magneticphases}}

\begin{figure}[t!]
\centering
\includegraphics[scale=1.0]{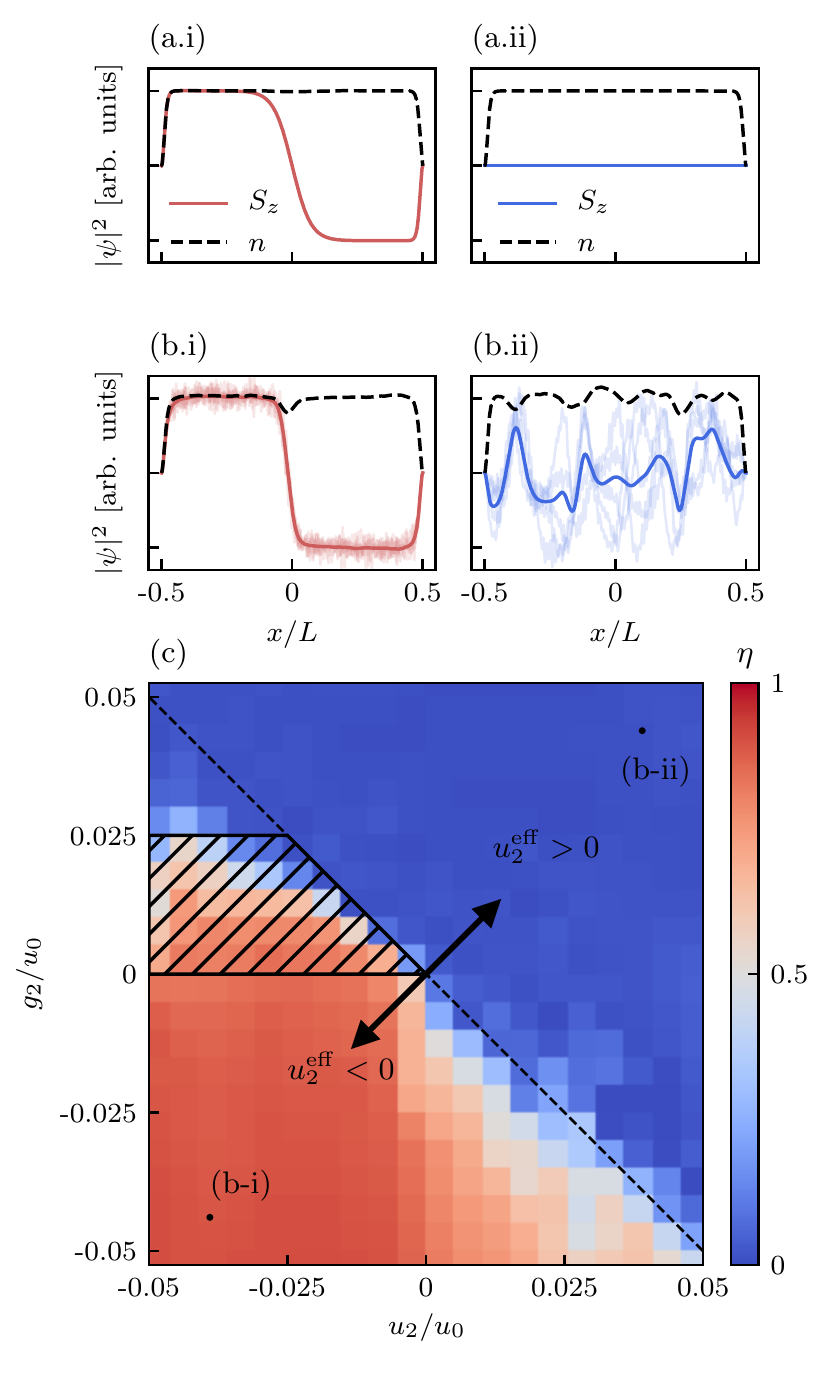}
\caption{(a) Ground state density (black, dashed curve) and spin density (solid curve) for (a.i) $u_2 < 0$ and (a.ii) $u_2 > 0$ . (b)  Steady state density (black, dashed curve) and spin density (solid curve) for (b.i) $u^{\rm eff}_2 \lesssim 0$ and (b.ii) $u^{\rm eff}_2 \gtrsim 0$, averaged over \SI{100}{\ms}. Semi-transparent curves indicate $S_z$ without time averaging. (c) Steady-state phase diagram as a function of $u_2/u_0$ and $g_2/u_0$ (defined in text), showing magnetically ordered, easy-axis ferromagnet (red/lower left) or spin-disordered paramagnet (blue/upper right) phases. The black dashed line indicates the expected phase boundary at $u_2^{\rm eff} = 0$, and the hatched region indicates bistability depending on the initial phase. The system enters an easy-axis ferromagnet if the initial condition is (a.i), and a spin-disordered paramagnet if the initial condition is (a.ii). \label{Fig:phasediagrammain}}
\end{figure}

We now focus on feedback-tuned spin-dependent interactions with $g_2\neq 0$ and $g_0=0$.  Guided by our toy model, we expect the steady-state phase diagram of a two-component BEC to resemble the ground state phase diagram for $u_2$. The ground state density $n(x)$ and spin density $S_z(x)$ are shown in in Fig.~\ref{Fig:phasediagrammain}~(a). For $u_2 >0$, the ground state is an easy-plane ferromagnet with $S_z(x) = 0$, while for $u_2 < 0$ the ground state is an easy-axis ferromagnet, consisting of spin-polarized domains~\cite{Trippenbach2000, Barnett2006, Kawaguchi2011, De2014}, separated by a domain wall.

Using the measurement and feedback procedure outlined in Sec.~\ref{SubSubSec:toy}, we apply a forcing potential
\begin{equation}
\check{V}_{\rm f}(x, t) = g_2\e_z(x, t)\check{\sigma}^z, 
\label{Eqn:Vf1}
\end{equation}
along with a cooling potential $\check{V}_{\rm c}$, to be described in Sec.~\ref{Sec:cooling}. 
Equation~\eqref{Eqn:Vf1} changes the effective spin dependent interaction strength via the gain $g_2$, based the estimator of the spin density $\e_z$. The effective Hamiltonian for this protocol is 
\begin{equation}
\check{\hat{\mc{H}}}_{\rm eff} \approx \left[\hat{H}_0 + u_0n\right]\check{\mathbb{1}} + \check{V}_c + \left[u_2S_z + g_2\e_z\right]\check{\sigma}^z.
\label{Eqn:Heff1}
\end{equation}
The phase diagram is now a function of two variables: spin-dependent interaction strength $u_2$ and signal gain $g_2$, which give an effective interaction strength $u^{\rm eff}_2 \approx u_2 + g_2$. Examples of the two steady-state phases are shown in Fig.~\ref{Fig:phasediagrammain}~(b). Both phases have uniform density, but with very different spin character. For $u^{\rm eff} \lesssim 0$, the system is an easy-axis ferromagnet with well defined, spin polarized domains. For $ u^{\rm eff}_2 \gtrsim 0$ the system enters a spin-disordered paramagnetic phase, with large spin fluctuations. Fig.~\ref{Fig:phasediagrammain}~(b) shows the spin density averaged over \SI{100}{\ms} (darker solid curve) and ten individual time traces (semi-transparent curves). The individual time traces show that the spin is essentially static in the ferromagnetic phase but has large spatiotemporal fluctuations in the paramagnetic phase.

Figure~\ref{Fig:phasediagrammain}~(c) shows the steady-state phase diagram as a function of $u_2/u_0$ and $g_2/u_0$. As expected, the phase diagram is divided into two regimes delineated by  $u_2^{\rm eff} = 0$ (black dashed curve). We quantify the steady-state phase using a time-separated correlation function of magnetization,  
\begin{equation}
\eta = \frac{1}{\mathcal{A}}\int d\tau \int dt~dx~ \frac{S_z(t + \tau, x)S_z(t, x)}{n(t + \tau, x)n(t, x)}, 
\label{Eqn:Eta}
\end{equation}
where $\mathcal{A}$ is an overall normalization factor. A condensate with well defined domains gives $\eta \gtrsim 0.5$; for the ground state with a single domain wall $\eta \approx 1$. The disordered paramagnet phase with fluctuating magnetization has $\eta \approx 0$, because the local magnetization at any point $x$ fluctuates strongly in time. 

Like many magnetic systems, this system exhibits hysteretic behavior. When $g_2 < 0$, the easy-axis phase is robust to the initial condition of the system and over many different repetitions of the simulation with different noise realizations. The phase in the region where $u^{\rm eff} \lesssim 0$ with $u_2 < 0$ and $g_2 > 0$ is sensitive to the initial state, denoted by the hatched region in Fig.~\ref{Fig:phasediagrammain}~(c). In this region, the steady state of the system is an easy-axis ferromagnet only if it was initially in the ferromagnetic ground state with $u_2 < 0$, as in \ref{Fig:phasediagrammain}~(a.i). For the easy-plane ground state, as in Fig.~\ref{Fig:phasediagrammain}~(a.ii), domains do not form. We discuss this steady-state behavior for the easy-plane initial condition in Appendix \ref{App:EasyPlane}.

In the following sections we examine the robustness of the feedback-induced magnetic phases and feedback cooling. We show that despite repeated weak measurements and feedback the condensate remains largely intact over the $\sim$ \SI{4}{\s} time period of the simulation. Furthermore, by changing the effective interaction via feedback we demonstrate tunability between different steady-state phases. Spatially-resolved, time dependent feedback therefore provides a tool to dynamically change effective interactions in cold atom systems. 		

\section{Feedback Cooling~\label{Sec:cooling}}

% !TEX root = main.tex
% Informs TeXShop to look for the main file. 

Measurement backaction adds excitations to the condensate. The aim of feedback cooling is to apply feedback using information from the measurement signal to suppress the excitations, thereby stabilizing the condensate and preventing runaway heating. In this section we develop a feedback cooling protocol for single and multicomponent condensates which ensures the stability of the condensate during measurement and feedback. We connect the continuous measurement limit presented in Sec.~\ref{SubSec:Formalism} to the experimental reality of discrete measurements. We then develop a feedback cooling protocol using a single discrete measurement as a building block. Finally, we show that during this protocol the condensate fraction and entropy reaches a steady-state, but the GPE energy functional continues to slowly increase.

\subsection{Single Measurement Protocol\label{SubSec:SingleMsrmt}}
The continuous measurement limit is typically assumed \emph{a priori} by taking $\dt \rightarrow 0$. Since the variance of the measurement signal in Eq.~\eqref{Eqn:Mresult} is $\propto 1/dt$, the variance in the measurement record diverges in this limit. However, no physical measurement is infinitely fast. Integrating Eq.~\eqref{Eqn:Mresult} over a small time window therefore yields a `single measurement'. By considering this type of measurement, we can quantify a measurement protocol which extracts maximal information from the condensate while minimizing the negative effects of backaction. As in Sec.~\ref{SubSubSec:toy}, here we consider measurements of a single component condensate and drop the $\s$ index. It is straightforward to generalize this procedure to multicomponent condensates.

Consider a time-integrated version of Eq.~\eqref{Eqn:Mresult} over an interval $\Delta t$, giving a single measurement of density. The measurement result is $\mathcal{M}(x) = n(x) + \bar{m}(x)/\kappa$, where the measurement strength $\kappa =\sqrt{\Delta t}\p$. The spatial quantum projection noise is $\bar{m}(x)$ where $\tilde{\bar{m}}_{k}$ has the same Fourier space statistics previously discussed, with $\overline{\tilde{\bar{m}}_{k}} = 0$ and $\overline{\tilde{\bar{m}}_{k}\tilde{\bar{m}}_{k'}} = L\delta_{kk'}\Theta(|k|-k_{\rm c})/2$. Directly after measurement, the updated wavefunction is $\psi_{|\mathrm{M}}(x)\approx \psi(x) + \kappa \bar{m}(x)\psi(x)$. Thus, there exists an optimal measurement strength 
\begin{equation}
\kappa_{*} \approx \sqrt{\frac{1}{2~\mathrm{max}[n(x)]}}, 
\label{Eqn:idealcoupling}
\end{equation}
such that the measurement outcome matches the post-measurement density $n_{|\rm M}$ exactly, i.e. $\mc{M}(x) = n_{|\mathrm{M}}(x)$. In principle, the optimal measurement strength depends on the local density, however as this is difficult to implement experimentally we instead approximate $\kappa_{*}$ to be constant. We then use this coupling value for feedback cooling. 

If we could find a potential $V_{\mathrm{c}|\mathrm{M}}(x)$ for which the post measurement state is the ground state, $\psi_{|\mathrm{M}}(x)$ would satisfy the stationary GPE
\begin{equation}
\mu \psi_{|\mathrm{M}} = \left[\hat{H}_0 + u_0n_{|\rm M} + V_{\mathrm{c}|\mathrm{M}}\right]\psi_{|\mathrm{M}}.
\label{Eqn:GPEstationary}
\end{equation}
In our feedback cooling protocol, we first apply the potential $V_{\rm c|\mathrm{M}}(x)$ for which the post-measurement state \emph{would} be the ground state (assuming a uniform phase). Then we approach the initial state by slowly -- adiabatically -- ramping off the applied cooling potential. We approximate $V_{\mathrm{c}|\mathrm{M}}$ using the Thomas-Fermi (TF) approximation of Eq.~\eqref{Eqn:GPEstationary}, giving $V_{\mathrm{c}|\mathrm{M}}(x) = \mu -  u_0n_{|\rm M}(x)$. We then make the substitution $u_0n_{|\rm M}(x) \rightarrow g_{\rm c}\mc{M}(x)$ where $g_{\rm c}$ is the cooling gain, an externally adjustable parameter (for which the expected value of $u_0$ is found to be optimal). This gives the feedback cooling potential function
\begin{equation}
V_{\mathrm{c}|\mathrm{M}}(x, t) = \left[\mu -  g_{\rm c}\mc{M}_{t_{\rm m}}(x)\right] f\left(t-t_{\rm m}\right), 
\label{Eqn:singlemeasurementpotential}
\end{equation}
where $t_{\rm m}$ is the time of the measurement and $f(t)$ is a ramp off function where $f(0) = 1$ and $f(t\rightarrow \infty) = 0$. In practice we use $f(t-t_{\rm m}) \approx 1-\gamma(t-t_{\rm m})$ where $\ramprate$ is the ramp-off rate.

\subsection{Bogoliubov Theory for Single Measurement Protocol\label{SubSec:BdG}}

Here we provide an analytical solution of the single-measurement-feedback protocol described above using Bogoliubov theory~\cite{Pitaevskii2003}, with periodic boundary conditions. After making the Bogoliubov transformation, small excitations above the ground state of a weakly interacting spinless BEC with density $n$ are described by the Hamiltonian
\begin{equation}
    \hat{H}_{\rm ph} = \sum_k \epsilon_k \hat{b}^\dagger_k\hat{b}_k,
\end{equation}
where $\hat{b}^\dagger_k$ describes the creation of a Bogoliubov phonon with momentum $k$ and energy $\epsilon_k = \mu\xi |k| \sqrt{\xi^2 k^2 +2 }$. To facilitate our analytic treatment, we focus on the weak measurement regime, in which at most one phonon mode is occupied, leading to wavefunctions of the form $|\psi\rangle = \alpha|\text{vac}\rangle + \sum_k \beta_k |k\rangle$, where  $|k\rangle=\hat{b}^\dagger_k |\text{vac}\rangle$, and $|\text{vac}\rangle$ is the phonon vacuum.

Measurement backaction is described by the Kraus operator
\begin{align}
\hat{K} = \exp\left\{-\frac{\kappa^2}{2} \int dx \left[\delta\hat{n}(x) - \frac{\bar{m}_{t_m}(x)}{\kappa}\right]^2\right\},
\end{align}
with the density difference operator $\delta \hat{n}(x) \equiv \hat{n}(x) - n$. In the phonon basis $\delta \hat{n}(x)$ can be expressed as a sum $\delta \hat{n}(x) = \sqrt{n/L}\sum_k (c_ke^{-ikx}\hat{b}_k+\text{h.c.})$ of phonon creation and annihilation operators, with $c_k= [1 +2/(\xi k)^2]^{-1/4}$.

In this representation, the feedback cooling operator derived from \eqref{Eqn:singlemeasurementpotential} is
\begin{equation}
   \hat{V}_{\mathrm{c}|\mathrm{M}}(t) = \int dx~ V_{\mathrm{c}|\mathrm{M}}(x, t)  \hat{n}(x).
\end{equation}
Assuming adiabatic evolution, with ramp-off rate $\gamma \to 0$, and using first order perturbation theory, the operator describing the cooling protocol is
\begin{align}
    \hat{R}_{|\rm m} 
    &= 1+\sum_{k}  \frac{g_{\rm c} c_k\sqrt{n}}{\kappa\epsilon_k\sqrt{L}}\left[\tilde{\bar{m}}_{t_{\rm m}}(k)  \hat{b}_k - \text{h.c.}\right]. 
\end{align}
This expression is valid for $g_{\rm c} c_k\sqrt{n} \ll \kappa\epsilon_k\sqrt{L}$. The probability of finding a phonon in state $|k\rangle$  after a measurement-feedback cycle is
\begin{equation}
    \bar{P}_{k} = \overline{|\langle k | \hat{R}_{|\rm m} \hat{K} |\text{vac}\rangle|^2} = \frac{n\kappa^2 c_k^2}{2}\left(1 -\frac{g_{\rm c}}{\kappa^2\epsilon_k}\right)^2 \Theta(|k|-k_{\rm c}).
    \label{Eqn:probability}
\end{equation}

\begin{figure}[t!]
\centering
\includegraphics[scale=1.0]{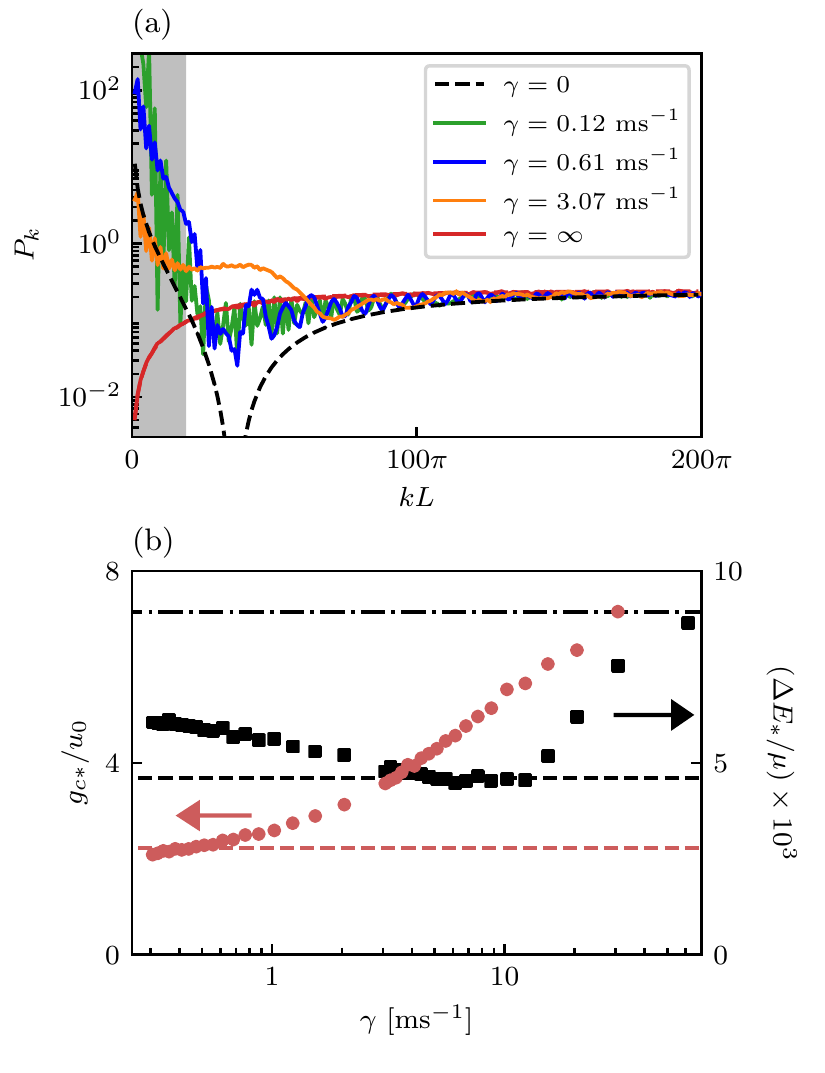}
\caption{Comparison between Bogoliubov theory and stochastic GPE simulation for a single measurement-feedback cycle for a system initially in the ground state. (a) Phonon population.  Black, green, blue, orange, and red curves indicate $\gamma = \SI{0}{\ms}^{-1}$, $\gamma =\SI{0.12}{\ms}^{-1}$, $\gamma =\SI{0.61}{\ms}^{-1}$, $\gamma =\SI{3.07}{\ms}^{-1}$ and $\gamma =\infty$.  Dashed curves result from Bogoliubov theory [Eq.~\eqref{Eqn:probability} with $g_{\rm c} =u_0$ and $g_{\rm c} =0$, corresponding to $\gamma = 0$ and $\gamma = \infty$ respectively], while solid curves derive from GPE simulations ($3000$ trajectories).   The Bogoliubov and GPE results coincide for $\gamma =\infty$ (red).  The grey region marks wavenumbers for which first order perturbation theory fails. (b) Gain $g_{\rm c*}$ (red circles) for which the energy increase $\Delta E_*$ (black squares) is minimized, plotted as a function of $\gamma$. For each point, we fit Eq.~\eqref{Eqn:energyincrease} to the GPE simulation result with $A, g_{\rm c*}$, and $\Delta E_*$ as free parameters. Horizontal dashed lines indicate the Bogoliubov prediction of $\Delta E_*$ and $g_{\rm c*}$, and dash-dotted line shows energy increase without feedback cooling (i.e., $\gamma =\infty$). \label{Fig:bdgcompare}}
\end{figure}

We draw two conclusions from this result: (1) Setting $g_{\rm c}=0$ gives the probability $n\kappa^2 c_k^2/2$ that the measurement created a phonon in state $|k\rangle$; and (2) the phonon mode with energy $\epsilon_{k, \text{opt}} = g_{\rm c} \kappa^{-2}$ can be perfectly cooled with this protocol. Figure~\ref{Fig:bdgcompare}~(a) compares Eq.~\eqref{Eqn:probability} with our stochastic GPE simulation with a linear ramp-off function $f(t)$. The analytic calculation exactly reproduces the numerically predicted phonon distribution immediately following a single measurement (red curve), while the results with cooling have additional periodic features resulting from the finite ramp-off rates in the simulations. The shaded region denotes the parameters for which our perturbation theory is inapplicable.

In the thermodynamic limit $L\gg\xi$, the per-particle energy after one measurement-feedback cycle
\begin{align}
    \Delta E = \frac{1}{2\pi n} \int dk~  \epsilon_k \bar{P}_{k} =A(g_{\rm c}-g_{\rm c*})^2+ \Delta E_*
    \label{Eqn:energyincrease}
\end{align}
is parabolic. With $\xi\gg1/k_{\rm c}$, the minimal per-particle energy increase $ \Delta E_*/\mu = \kappa ^2  \phi_{\rm c}^2(\pi  \phi_{\rm c} - 6 \sqrt{2})/(6 \pi^2 \xi)$ occurs for a gain
\begin{align}
\frac{g_{\rm c*}}{u_0} &=\frac{2 \sqrt{2} \kappa ^2  n \phi_{\rm c} }{\pi },
    \label{Eqn:optimalgain}
\end{align}
where  $\phi_{\rm c} =  k_{\rm c}\xi/\sqrt{2}$ parameterizes the cutoff and $A=(4 \sqrt{2} \kappa ^2 \mu  \xi)^{-1}$.

Figure \ref{Fig:bdgcompare}~(b) compares the optimal energy increase predicted by Eq.~\eqref{Eqn:energyincrease}, with that obtained from numerical simulations of the stochastic GPE (horizontal black dashed line and black squares, respectively), and the corresponding optimal gains are denoted by the red circles. The GPE simulation exhibits three regimes: (1) For very rapid ramps $\gamma\to \infty$, the adiabatic assumption is invalid, and the GPE optimal gain is larger than anticipated from analytic model. (2) In the adiabatic ramping regime where $\gamma\to 0$, we find both $g_{\rm c*}$ and $\Delta E_*$ converge, with $\Delta E_*$ greater than our predicted value. This results from phonon-phonon scattering processes redistributing phonons between modes, which is not included in our Bogoliubov theory. And, (3) in the intermediate regime ($\gamma$ between $\SI{3}{\ms}^{-1}$ and $\SI{10}{\ms}^{-1}$) our theory performs optimally and $\Delta E_*$ coincides with the analytic prediction, albeit with much higher gain. We note that the optimal gain $g_{\rm c} = u_0$ obtained in Sect.~\ref{SubSec:SingleMsrmt} is close to that predicted by Eq.~\eqref{Eqn:optimalgain}, where for the parameters in Fig.~\ref{Fig:bdgcompare}, $g_{\rm c*} \approx 2.8 u_0$.

\subsection{Continuous Feedback Cooling Protocol \label{SubSec:continuouscooling}}

\begin{figure}[t!]
\centering
\includegraphics[scale=1.0]{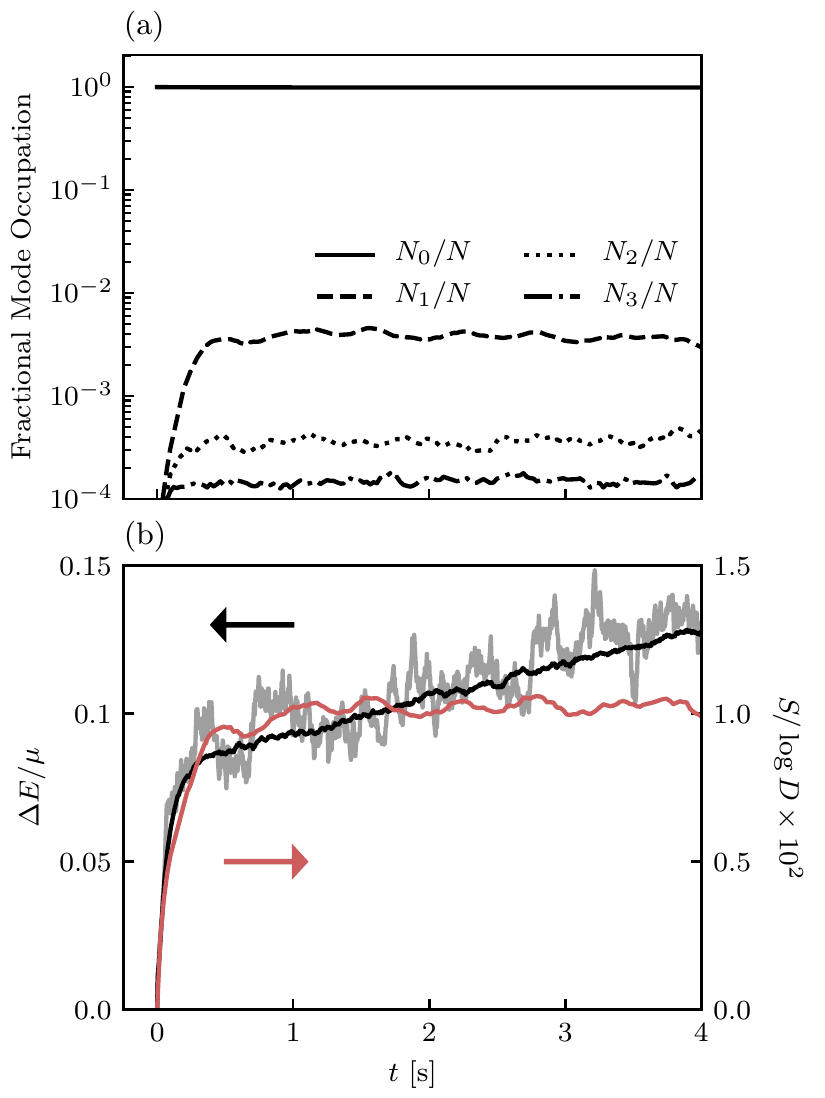}
\caption{Properties of single a component condensate under feedback cooling with gain $g_{\rm c} = u_0$, measurement strength $\kappa_* = 2.2\times10^{-3}$. Statistical properties were calculated from 128 independent stochastic trajectories.~(a)~Fractional occupation of the first four modes in the single-particle density matrix. The condensate fraction (solid curve) is $\approx 0.99$ in quasi-steady state.~(b)~Von Neumann entropy (red/light gray) and average energy (black) of the condensate. The gray curve is the energy for a single trajectory. \label{Fig:spinlesscooling} }
\end{figure}

The single measurement procedure described in Sec.~\ref{SubSec:SingleMsrmt} is a building block for continuous feedback cooling. We periodically measure the condensate with measurement strength $\kappa = \kappa_*\sqrt{\Delta t/\tau}$ where $\kappa_*$ is the ideal single measurement strength in Eq.~\eqref{Eqn:idealcoupling} and $\tau$ is the filtering time constant for the measurement signal. The cooling potential is derived from the density estimator $\e(x, t)$~\cite{epsnote} and is decreased between measurements, as described by Eq.~\eqref{Eqn:singlemeasurementpotential}.

The effect of the cooling potential is to drive $\psi(x)$ toward it's ground state between measurements. This procedure leverages the optimal single measurement strength and signal filtering to measure the condensate more weakly. We implement this protocol numerically and simulate condensate evolution under measurement and feedback using Eqs.~\eqref{Eqn:dpsiH}-\eqref{Eqn:dpsif}. 

Here we simulate an elongated condensate with $N=10^5$ particles, healing length $\xi = $ \SI{0.8}{\um} and total system size $L = $ \SI{80}{\um}, computed for $k_{\rm c} = 2\pi/\lambda$ with $\lambda =$ \SI{780}{\nm}. The interval between measurements is set to $\dt = $ \SI{200}{\us} to match typical image acquisition times in experiment, and the estimator time constant and cooling ramp-off rate were set to $\tau = 1/\ramprate = $ \SI{4.6}{\ms}. We characterize the quasi-steady state by three metrics: condensate fraction, Von Neumann entropy, and energy, and find that the condensate remains remarkably coherent throughout the feedback cooling protocol. Upon implementing continuous feedback cooling, the condensate fraction and Von Neumann entropy reach a steady state while the GPE energy functional slowly increases, as shown in Fig.~\ref{Fig:spinlesscooling}. 

We calculate the condensate fraction using the Penrose-Onsager criteria~\cite{Penrose1956}. Per this criteria, upon diagonalizing  the one body density matrix $\hat{\rho}$ as $\hat{\rho}|n\rangle = N_n|n\rangle$, a condensate is present in mode $|n\rangle$ if it's eigenvalue is $N_n \sim \mathcal{O}(N)$ where $N$ is the total number of particles. We obtain $\hat{\rho}$ from an ensemble of stochastic trajectories of pure states~\cite{Daley2014}, starting from the GPE ground state. In Fig.~\ref{Fig:spinlesscooling}~(a) we show the four largest eigenvalues of $\hat{\rho}$, normalized by $N$, giving a measure of the fractional occupation in each mode. The condensate fraction is the largest eigenvalue, which stabilizes at $\approx 0.99$, with a secondary mode having an occupation fraction of $\approx 0.01$. The remaining eigenvalues are orders of magnitude smaller than the leading two; therefore those modes have negligible occupation.

The second metric we use to characterize the steady state is the Von Neumann entropy, defined as $S = \mathrm{Tr}\left[\hat{\rho}\log\hat{\rho}\right]$. As shown in Fig.~\ref{Fig:spinlesscooling}~(b), $S$ saturates at $\approx 0.01$ of it's maximum possible value $\log(D)$, where $D$ is the Hilbert space dimension. This is consistent with the final condensate fraction of $\approx 0.99$. We extract an equilibration time $\tau_{\rm eq} \approx $ \SI{200}{\ms} by fitting $S$ to the function $S(t) \approx S_0 (1-e^{-t/\tau_{\rm eq}})$. 

The third metric, energy, does not reach a constant value, rather it slowly increases even after the condensate fraction and entropy saturate, as shown in Fig.~\ref{Fig:spinlesscooling}~(b). Here we define energy in terms of the per-particle GPE energy without any feedback terms present. The final energy after \SI{4}{\s} of evolution is $\sim 0.15~\mu$, indicating a $15\%$ increase from the ground state value throughout the protocol. We determined that this energy increase is due to the gradual population of modes above the momentum cutoff which cannot be directly addressed by feedback cooling. However, this increase is slow enough to provide ample time (on the order of seconds) for additional experiments while the condensate is being measured. 

Cooling for the two-component case proceeds similarly, but with cooling applied in the spin and density channels separately. Weak measurements add magnons (spin waves) in addition to phonons~\cite{Stamper2013}. For the easy-axis ground state with $u_2 < 0$, the results are qualitatively the same as as the single component case, with the final condensate fraction reduced to $\approx 0.85$, indicating cooling is not quite as efficient for the two-component system. However, in the easy-plane case (i.e. $u_2 >0$), cooling is not as effective at long times and the condensate enters a spin-disordered phase with large spin fluctuations and a lower condensate fraction of $\approx 0.35$. The cooling protocol for two-component condensates is discussed in Appendix~\ref{App:SpinorCooling}.

\section{Feedback Induced Magnetic Phases~\label{Sec:spinfeedback}}

% !TEX root = main.tex
% Informs TeXShop to look for the main file. 

In this section, we elaborate on the steady state magnetic phases and their measurement signatures. The phase diagram in Fig.~\ref{Fig:phasediagrammain}~(c) was computed for a gas of $N = 10^5$ $^{87}$Rb atoms with healing length $\xi = $ \SI{0.8}{\um} and total system length $L = $ \SI{80}{\um}, with feedback both to control the effective interactions and cool the system. In all of our simulations, feedback cooling is continuously applied. We add the forcing feedback $\check{V}_{\rm f}(x, t)  = g_2\e_z(x, t)\check{\sigma}^z$ in the time window from \SI{1}{\s} to \SI{3}{\s} and allow the simulations to continue until the total run time reaches \SI{4}{\s}.

Figure~\ref{Fig:phasediagrammain}~(c) shows that the magnetic phase of the system reaches a steady-state governed by the effective spin-dependent interaction strength $u_{2}^{\mathrm{eff}} = g_2 + u_2$ while the forcing potential is on, leading to the easy-axis ferromagnet and spin-disordered paramagnetic phases discussed in Sec.~\ref{SubSec:magneticphases}. The spin-dependent interaction strength $u_2$ and gain $g_2$ serve as tunable parameters. 

\begin{figure}[t!]
\centering
\includegraphics[scale=1.0]{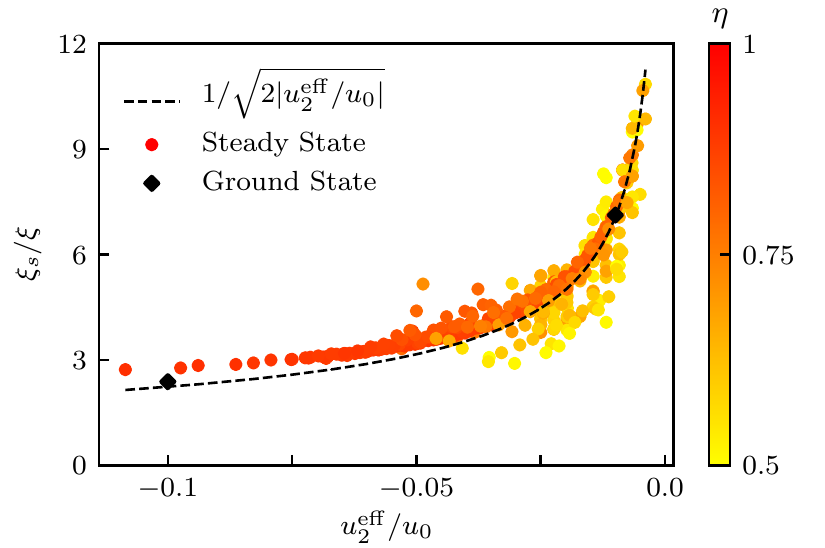}
\caption{Spin healing length as a function of effective spin-dependent interaction strength $u_{2}^{\mathrm{eff}} = g_2 + u_2$ for data shown in Fig.~\ref{Fig:phasediagrammain}~(c) phase diagram with $u_{2}^{\mathrm{eff}} < 0$. The colored markers indicate  the calculated spin healing length averaged over \SI{1.6}{\s} window.  The black markers indicate the spin healing length for a ground state system (i.e., no feedback) with $u_2$ equal to the marked value of $u_{2}^{\rm eff}$. The dashed curve indicates the predicted spin healing length $\xi_s = \xi/\sqrt{2|u_{2}^{\rm eff}/u_0|}$ with no fitting parameters.\label{Fig:SpinHealingLength}}
\end{figure}

The easy axis ferromagnetic phase for $u_{2}^{\mathrm{eff}} < 0$ exhibits well defined, spin-polarized domains. The order parameter $\eta$ for this phase is the time-separated correlation function of the magnetization, given in Eq.~\eqref{Eqn:Eta}. We find that $\eta \gtrsim 0.5$ indicates the existence of persistent domains. We can identify an effective spin healing length $\xi_{s} \propto 1/\sqrt{|u_{2}^{\rm eff}|}$ in this phase, similar to the spin healing length in closed two-component systems~\cite{De2014}. Changing $u_{2}^{\rm eff}$ via the feedback strength thus alters the spin healing length in the steady state. 

\begin{figure}[t!]
\centering
\includegraphics[scale=1.0]{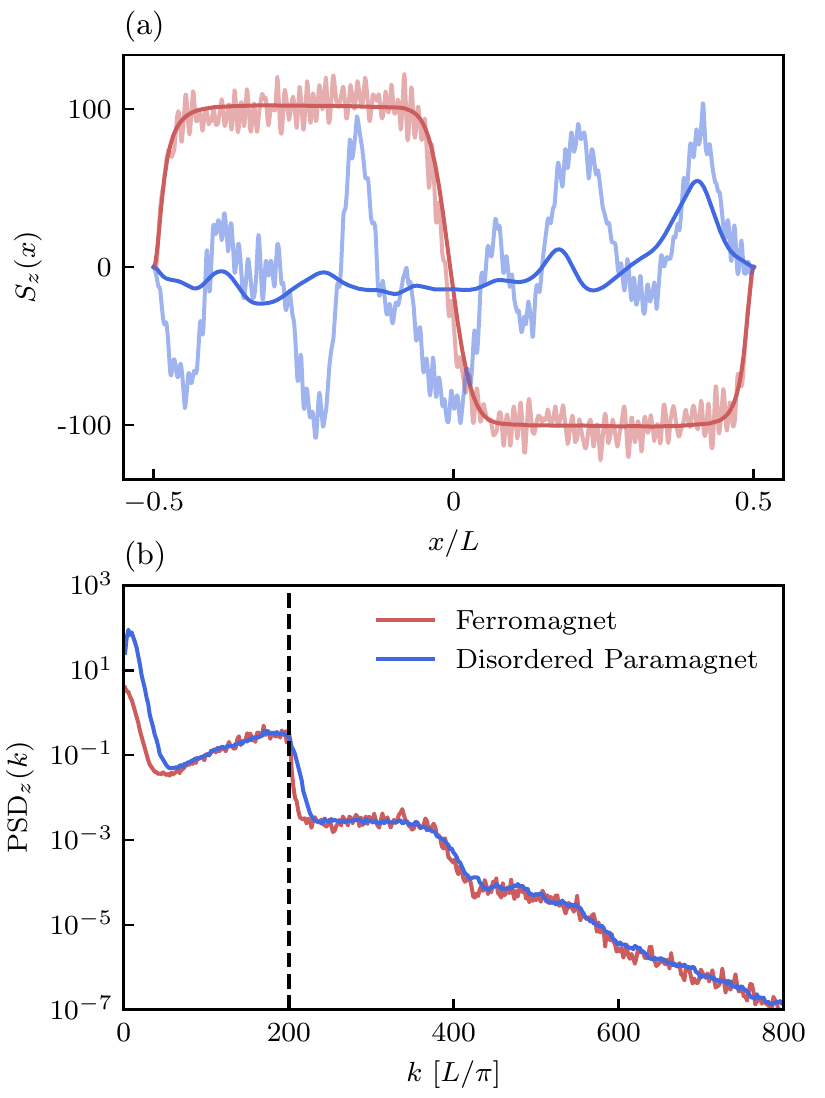}
\caption{~(a)~Real space spin density $S_z(x)$ computed in the ferromagnetic and disordered paramagnetic phase. The solid curve shows the time-averaged signal over \SI{1}{\s} and the semi-transparent curve indicates a single time trace.~(b)~The corresponding power spectral density of fluctuations in each phase. The vertical dashed line indicates the momentum cutoff $k_{\rm c}$.~\label{Fig:PSD}}
\end{figure}

Figure~\ref{Fig:SpinHealingLength} shows the effective spin healing length, obtained by fitting the spin density $S_z(x)$ to a function with $N_{\rm d}$ domains, where 
\begin{equation}
S_z(x) = \pm \mathcal{S}~\Pi_{n=1}^{N_{\rm d} - 1} \tanh\left(\frac{x-x_n}{\xi_{s}}\right).
\end{equation}
Here, $x_n$ are the positions of each domain wall, $\mathcal{S}$ is the overall amplitude of domains, and $\xi_{s}$ is the spin healing length. The $\pm$ sign in front accounts for the polarity of the domain signal (i.e. which domain is at the edge), as the measurement and feedback process spontaneously breaks a $\mathbb{Z}_2$ symmetry to determine the domain orientations~\cite{Hurst2019, Pintos2019}. 

The spin healing length diverges upon approaching the transition at $u_{2}^{\mathrm{eff}} = 0$, indicated system behavior that is analogous to the expected phase transition from changing the interaction parameters. The markers in Fig.~\ref{Fig:SpinHealingLength} are color-coded based on the value of the $\eta$, where we can see that for lower values there is more variability in the data. This is because lower values of $\eta$ generally correspond to a spin texture with multiple domains, where there is movement of the domain boundaries over time due to fluctuations parameterized by the nonzero enropy~\cite{Hurst2019}. The black diamonds in Fig.~\ref{Fig:SpinHealingLength} show the spin healing length obtained for the corresponding closed system ground state, and the dashed curve is the computed functional dependence $\xi_s = \xi[u_0/2|u_{2}^{\rm eff}|]^{1/2}$ for $u_2^{\rm eff} < 0$, which shows excellent agreement with the simulations.  

The disordered paramagnetic phase is characterized by a spatially and temporally fluctuating spin structure. An example of these fluctuations in real space is shown in Fig.~\ref{Fig:PSD}~(a). In the disordered paramagnetic phase, a spin healing length is not well defined. The power spectral density (PSD) of the spin, 
\begin{equation}
\mathrm{PSD}_z(k, t) = |\tilde{S}_z(k, t) - \bar{\tilde{S}}_z(k, t)|^2, 
\end{equation}
provides a measure of how much the spin fluctuates~\cite{De2014}. Here $\bar{S}_z(x)$ is the time-averaged value of the spin density and $\tilde{S}_z(k, t)$ is the Fourier transform of $S_z(x,t)$. 

Figure~\ref{Fig:PSD}~(a) shows $\mathrm{PSD}_z(k)$ in the steady-state magnetic phase averaged over \SI{1}{\s}. At low momenta the signature for the disordered phase is significantly higher than for the easy-axis ferromagnetic phase. The large fluctuations in spin are thus a signature of the paramagnetic phase which can be deduced from the measurement signals. Above the cutoff $k_{\rm c}\lambda = 200\pi$ indicated by the black, dashed line, we see additional spectral features at multiples of $k_{\rm c}$, indicating higher-order resonances due to the measurement process. Population of modes above the cutoff leads to a gradual increase in energy and affects cooling, as discussed in Sec.~\ref{SubSec:continuouscooling}.
 
\section{Outlook~\label{Sec:conclusion}}

% !TEX root = main.tex
% Informs TeXShop to look for the main file. 

Hamiltonian engineering for multicomponent Bose gases has been achieved at the level of the single-particle Hamiltonian via synthetic gauge fields~\cite{Lin2009, Goldman2014}, spin-orbit coupling~\cite{Lin2011, Galitski2013, Kroeze2019}, and spin-dependent potentials~\cite{Jimenez2012, Lu2016}. The ability to tune the character and strength of interactions beyond those already present in the system has heretofore been limited to using Feshbach resonances~\cite{Theis2004}, which typically change only one interaction constant at a time, or via coupling to an external cavity field~\cite{Ritsch2013, Landini2018, Kroeze2018}. In contrast, our feedback technique can simultaneously change all the spin-dependent effective interaction strengths \emph{in situ}: not possible with Feshbach resonances or cavity mediated interactions.

Our result shows that spatially local feedback control based on a record of weak measurements is a viable route toward engineering effecting interactions in quantum gases. We demonstrated that a dynamical steady state can be engineered in a two-component Bose-Einstein condensate where the magnetic phase is determined by the interplay of the intrinsic and feedback-induced interaction strengths. 

Going beyond previous works~\cite{Hush2013, Hurst2019}, we implemented a cooling scheme which avoids runaway heating of the condensate during the feedback process. Further optimization of the cooling protocol will be important for experimental implementation. For example, Eq.~\eqref{Eqn:probability} suggests that the $k$ dependent gain $g_c(k) = n\kappa^2 \epsilon_k$ would lead to near-perfect cooling for all momentum states. Actual measurements have limited resolution, detector inefficiencies, and technical noise, which could possibly be addressed by further optimizing the cooling protocol.

The feedback control method of engineering effective Hamiltonians is flexible and allows for the introduction of tailored, spatially dependent effective interaction terms. Future work could implement nonlocal or time-dependent interactions which have no analogue in closed systems. Our protocols can be generalized to higher dimensions, and could stabilize topological defects such as non-Abelian vortex anyons which are unstable in closed systems~\cite{Mawson2019}. Finally, our methods enable real-time feedback control, so over the course of one experiment we can study both quasi-steady-state behavior and dynamics.

\begin{acknowledgments}
This work was partially supported by NIST and NSF through the Physics Frontier Center at the JQI. HMH acknowledges the support of the NIST NRC postdoctoral program.
\end{acknowledgments}

\appendix 

% !TEX root = main.tex
% Informs TeXShop to look for the main file. 

\section{Simulation Parameters \label{App:Params}}

Here we briefly review the simulation method for Eqs.~\eqref{Eqn:dpsiH}-\eqref{Eqn:dpsif} and the parameters we use in this work. All simulations have $N= 10^5$ atoms and we consider a quasi-1D system of length $L =$ \SI{80}{\um} with hard wall boundary conditions such that $\Psi(x = -L/2) = \Psi(x = L/2) = 0$. Hard-wall boundaries can be implemented using flat-bottomed traps instead of a harmonic one~\cite{Meyrath2005}. The momentum cutoff is $k_{\rm c} = 2\pi/\lambda$ with $\lambda = $ \SI{780}{\nm} being the wavelength of imaging light. We simulate a single component condensate in order to study steady state behavior under feedback cooling in Sec.~\ref{Sec:cooling}. Elsewhere, we simulate a two-component condensate with an easy-axis magnetic ground state, i.e. $u_2 < 0$, or easy-plane ground state with $u_2 > 0$. In the main text results are presented using the easy-axis ground state with $u_2 = 0.01u_0$ as the initial condition. 

The system is initialized in it's ground state by solving the GPE in imaginary time. The natural units for this set up are the total system length $L$ and the chemical potential $\mu = \hbar^2/2m\xi^2$ as the unit of energy where $\xi = $ \SI{0.8}{\um} is the healing length. Upon re-scaling the variables to unitless quantities $x \rightarrow xL$, $t \rightarrow t(2m_{\rm a}\xi^2/\hbar)$, $\psi_{\up(\down)} \rightarrow \sqrt{N/L} \psi_{\up (\down)}$, the Hamiltonian in Eq.~\eqref{Eqn:GPEstationary} is 

\begin{equation}
\check{\hat{\mathcal{H}}} = \left[-\frac{\xi^2}{L^2}\frac{\partial^2}{\partial x^2} + n(x)\right]\check{\mathbb{1}} + \frac{u_2}{u_0}S_z(x)\check{\sigma}^z,
\label{Eqn:H0unitless}
\end{equation}
where $\int dx~n(x) = 1$. Therefore, the spinless case has one free parameter $\xi/L$ and the two-component case has the additional free parameter $u_2/u_0$. For our parameters we have $\xi/L = 0.01$ and we consider different values of $u_2$. We simulate the nonlinear dynamics using a second-order symplectic integration method~\cite{Symes2016}. In these units it is natural to express $u_2$ and the gain strengths $g_0, g_2$, etc in units of $u_0$. 

In order to simulate a small measurement interval (approaching the continuous measurement limit), we consider a separation of timescales $\dt \ll \tau$ such that the measurement interval $\dt$ of the system is much shorter than the signal filtering timescale $\tau$ for any observable. This enables us to write the evolution Eq.~\eqref{Eqn:dpsiH}-\eqref{Eqn:dpsif} as continuous time stochastic differential equations. 

\begin{figure}[t!]
\centering
\includegraphics[scale=1.0]{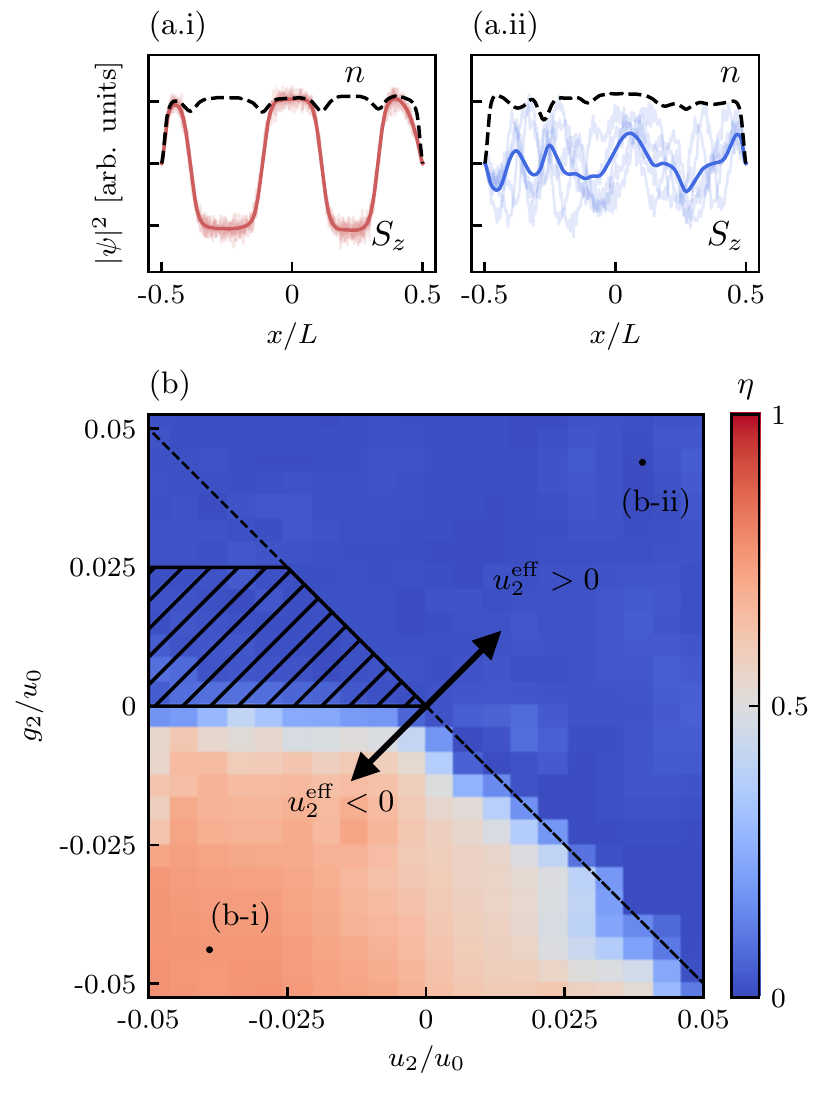}
\caption{(a)  Steady state density (black, dashed curve) and spin density (solid curve) for (a.i) $u^{\rm eff}_2 \lesssim 0$ and (a.ii) $u^{\rm eff}_2 \gtrsim 0$, averaged over \SI{100}{\ms}. Semi-transparent curves indicate $S_z$ without time averaging. (b) Steady-state phase diagram as a function of $u_2/u_0$ and $g_2/u_0$ (defined in text), showing magnetically ordered, easy-axis ferromagnet (red/lower left) or spin-disordered paramagnet (blue/upper right) phases. The black dashed line indicates the expected phase boundary at $u_2^{\rm eff} = 0$, and the hatched region indicates bistability depending on the initial phase.\label{Fig:phasediagramappendix}}
\end{figure}

\section{Steady-State Phase Diagram for Easy Plane Initial Condition \label{App:EasyPlane}}
As indicated by the hatched region in Fig.~\ref{Fig:phasediagrammain}~(c)~, the steady state phase diagram has a region of bistability depending on the initial state of the system. In this Appendix we present the results for the phase diagram calculated using  the easy-plane ground state as the initial condition, shown in Fig.~\ref{Fig:phasediagramappendix}. In the steady-state magnetic phase, the system forms domains for $u_{2}^{\rm eff} < 0$ and $g_2 <0$. An example of the density and spin density in this region is shown in Fig.~\ref{Fig:phasediagramappendix}~(a.i), where we see that there are multiple domains in the spin texture. This is in contrast to the case presented in the main text where there is only one domain, due to the single-domain being the ground state. The number of domains depends on many parameters including $u_2$, $g_2$, and the timescale over which feedback is turned on. We consider further investigation of these variables to be outside the scope of this work. 

Unlike the easy-axis initial condition, the spin-disordered phase occurs for a wider range of parameters, most notably in the hatched region where $u_{2}^{\rm eff} = 0$ but $g_2 > 0$. The spin texture in this regime is shown in Fig.~\ref{Fig:phasediagramappendix}~(a.ii), which indicates relatively uniform density but a highly fluctuating spin texture. We suspect that the observed bistability could be due in part to the underlying cooling protocol for the two-component system, which can also affect the spin texture, as discussed in Appendix~\ref{App:SpinorCooling}.

\section{Two-Component Feedback Cooling\label{App:SpinorCooling}}

The density is measured in each component $\s$ with strength $\kappa = \kappa_*\sqrt{\Delta t/\tau_n}$ where $\Delta t$ is the measurement duration and $\tau_n$ is the low-pass filtering time constant for the total density. Measurements $\mathcal{M}_\up$ and $\mathcal{M}_\down$ are then combined to give a measurement of total density ($\mathcal{M}_\up$ + $\mathcal{M}_\down$) or spin density ($\mathcal{M}_\up$ - $\mathcal{M}_\down$), which is used in a low-pass filter to calculate the estimators $\e_n$ and $\e_z$. Crucially, the filtering works best when $\e_n$ and $\e_z$ have different filtering time constants; we use $\tau_n = $ \SI{4.6}{\ms} and $\tau_z = $ \SI{46}{\ms}, respectively. This is due to the different types of excitations in the two-component case, which can be phonons or magnons. Phonons have faster time dynamics than magnons, which necessitates different time constants in each channel. 

The spin-dependent cooling potential is 
\begin{equation}
\check{V}_{\rm c}(x, t) = V_{\mathrm{c},n}\left[\e_n, t\right]\check{\mathbb{1}} + V_{c,z}\left[\e_z, t\right]\check{\sigma}^z. 
\end{equation}
As in the spinless case, the potentials $V_{\mathrm{c},n}$ and $V_{c,z}$ are calculated after each measurement and then exponentially ramped off between measurements. Cooling in the density channel is done via the potential 
\begin{equation}
V_{\mathrm {c}, n}(x, t) = \left[\mu - g_{\rm c}\e_n(x, t)\right]e^{-\gamma_n(t-t_{\rm m})} \label{Eqn:spinorVcn}
\end{equation}
where $g_{\rm c}$ is the gain. This potential drives the total density toward a uniform state based on estimator $\e_n$ with ramp-off rate $\gamma_n$. Cooling for the spin sector is via the spin-dependent potential 
\begin{equation}
V_{\mathrm {c}, z}(x, t)  = g_{{\rm c}, z}\left[\bar{\e}_z(x, t) - \e_z(x, t)\right]e^{-\gamma_z(t-t_{\rm m})}\label{Eqn:spnorVcz}, 
\end{equation} 
where $\ramprate_z$ is the spin ramp-off rate, $g_{\mathrm{c}, z}$ is the cooling gain for the spin sector, and $\bar{\e}_z$ indicates a running time average of $\e_z$. This potential drives the spin density $S_z(x)$ toward it's time-averaged value, effectively cooling short wavelength (high momentum) spin fluctuations but allowing long-wavelength spin textures such as domain walls to remain intact. In practice we use $\ramprate_n^{-1} = \tau_n$ and $\ramprate_z^{-1} = \tau_z$, with the other parameters the same as for the spinless case. We calculate $\bar{\e}_z$ by averaging the original signal over a \SI{120}{\ms} time window. Cooling is most effective when the gain parameters are $g = u_0$ and $g_{\mathrm{c}, z} = u_2$. 

\begin{figure}[t!]
\centering
\includegraphics[scale=1.0]{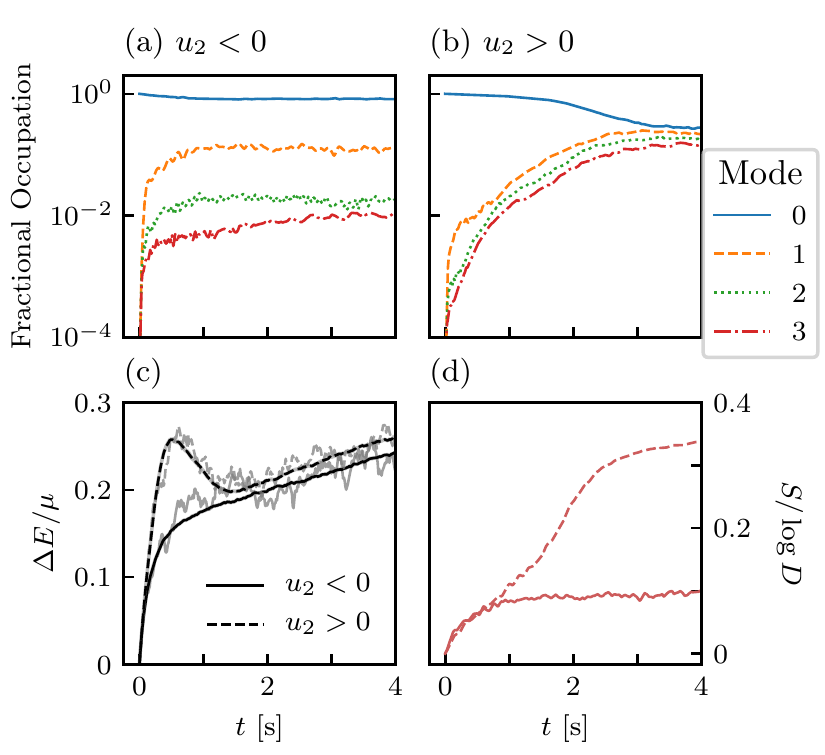}
\caption{Properties of a two component condensate under measurement and feedback cooling.~(a,b)~Fractional occupation of first four modes in the single-particle density matrix for~(a)~$u_2 < 0$ and ~(b)~$u_2 > 0$. The eigenvalue of the four highest-occupied modes is pictured. The condensate fraction (solid curve) is $\approx 0.85$ in the steady state for $u_2 < 0$ and $\approx 0.35$ for $u_2 > 0$.~(c)~Average energy (black) for a condensate with $u_2 < 0$ (solid curve) and $u_2 > 0$ (dashed curve) calculated from 124 independent stochastic trajectories. As in the spinless case, energy computed from the GPE energy functional increases slowly~(d)~Von Neumann entropy for a condensate with $u_2 < 0$ (solid curve) and $u_2 > 0$ (dashed curve).\label{Fig:spinorcooling}}
\end{figure}

As in the spinless case, feedback cooling drives the two-component condensate to a quasi-steady state. Condensate fraction and Von Neumann entropy  stabilize around constant values and the energy per particle increases slowly over the course of the simulation. We compute the energy from the GPE energy functional without any feedback terms present. The steady-state properties for cooling a two-component condensate are presented in Fig.~\ref{Fig:spinorcooling}. The results are qualitatively different for the case with $u_2 < 0$ (easy-axis ground state) and $u_2 > 0$ (easy plane ground state). 

The easy-axis case is similar to the spinless cooling results presented in the main text. In Fig.~\ref{Fig:spinorcooling}~(a) we present the condensate fraction for $u_2 < 0$, which can also be calculated for multicomponent condensates~\cite{Mason2014}. The condensate fraction is $\approx 0.85$ in the steady state with one additional mode having occupation $\approx 0.15$ and other modes having negligible occupation. The energy increase, shown in Fig.~\ref{Fig:spinorcooling}~(c) is $\approx 0.25\mu$. The Von Neumann entropy, shown in Fig.~\ref{Fig:spinorcooling}~(d) (solid curve) increases to about $10
\% $ of it's maximum value. These metrics indicate that the cooling protocol is effective for two-component condensates with $u_2 < 0$. Furthermore, we find that at the end of the cooling protocol the domain wall is still intact, showing that this spin dependent cooling protocol is effective both at maintaining a high level of condensation and preserving the spin structure. The equilibration time extracted from the entropy is $\tau_{\rm eq} \approx \SI{400}{\ms}$.

In the case of an easy-plane initial condition (i.e. $u_2 >0$), the cooling protocol is not as effective. In Fig.~\ref{Fig:spinorcooling}~(b)~we show the fractional occupation of the first four modes from the one-body density matrix. The condensate fraction (blue, solid curve) decreases to $\approx 0.35$ while the other modes also have fractional occupations of $\mathcal{O}(0.1)$. This indicates that the Penrose-Onsager criterion for condensation is violated in this regime. Furthermore, we find that the entropy $S$ increases considerably more than the easy-axis case, reaching a constant value of $\approx 0.4\log(D)$ after \SI{2}{\s} of time evolution. The entropy increase is likely being driven by an instability toward spin separation in the condensate. Under our current feedback protocol, the easy-plane ground state eventually enters a spin-disordered phase with large spin fluctuations, which accounts for the higher entropy and lower condensate fraction we observe. Future work could develop a feedback cooling protocol specifically for $u_2 > 0$ systems to combat this instability more effectively.

\bibliography{main}

\begin{thebibliography}{62}
\bibitem [{ano()}]{anote}%
  \BibitemOpen
  \href@noop {} {}\bibinfo {note} {In this work we use $\overline{~\cdot~}$ to
  denote a statistical average and $\langle\cdot\rangle$ to denote a
  quantum-mechanical expectation value. $\tilde{m}_{\s,k}$ indicates the
  Fourier transform of $m_\s(x)$.}\BibitemShut {Stop}%
\bibitem [{eps()}]{epsnote}%
  \BibitemOpen
  \href@noop {} {}\bibinfo {note} {We calculate the estimator $\e(x,t)$ after
  each measurement using a discretized version of
  Eq.~\eqref{Eqn:LowPass1}.}\BibitemShut {Stop}%
\end{thebibliography}
\end{document}